\documentclass[runningheads]{llncs}
\usepackage{graphicx}
\usepackage[T1]{fontenc}
\usepackage{listings}

\makeatletter
\def\@seccntformat#1{\@ifundefined{#1@cntformat}%
   {\csname the#1\endcsname\quad}  
   {\csname #1@cntformat\endcsname}
}
\let\oldappendix\appendix 
\renewcommand\appendix{%
    \oldappendix
    \newcommand{\section@cntformat}{\appendixname~\thesection\quad}
}
\makeatother

\usepackage[hidelinks]{hyperref}

\usepackage{color}

\usepackage{multirow}
\begin{document}
\title{Robots Still Outnumber Humans in Web Archives, But Less Than Before\thanks{This paper is an extended version of the paper accepted for publication at TPDL 2022}}

\author{Himarsha R. Jayanetti\inst{1}\orcidID{0000-0003-4748-9176} \and
Kritika Garg\inst{1}\orcidID{0000-0001-6498-7391} \and Sawood Alam\inst{2}\orcidID{0000-0002-8267-3326} \and Michael L. Nelson\inst{1}\orcidID{0000-0003-3749-8116} \and Michele C. Weigle\inst{1}\orcidID{0000-0002-2787-7166}}
\authorrunning{H. R. Jayanetti et al.}

\institute{Old Dominion University, Norfolk, VA 23529, USA \\ \email{\{hjaya002,kgarg001\}@odu.edu} \email{\{mln,mweigle\}@cs.odu.edu} \and
Wayback Machine, Internet Archive, San Francisco, CA 94118, USA
\email{sawood@archive.org}\\}
\maketitle
\begin{abstract}
To identify robots and humans and analyze their respective access patterns, we used the Internet Archive's (IA) Wayback Machine access logs from 2012 and 2019, as well as Arquivo.pt's (Portuguese Web Archive) access logs from 2019. We identified user sessions in the access logs and classified those sessions as human or robot based on their browsing behavior. To better understand how users navigate through the web archives, we evaluated these sessions to discover user access patterns. Based on the two archives and between the two years of IA access logs (2012 vs. 2019), we present a comparison of detected robots vs. humans and their user access patterns and temporal preferences. The total number of robots detected in IA 2012 is greater than in IA 2019 (21\% more in requests and 18\% more in sessions). Robots account for 98\% of requests (97\% of sessions) in Arquivo.pt (2019). We found that the robots are almost entirely limited to ``Dip’’ and ``Skim’’ access patterns in IA 2012, but exhibit all the patterns and their combinations in IA 2019.  Both humans and robots show a preference for web pages archived in the near past.

\keywords{ Web Archiving \and User Access Patterns \and Web Server Logs \and Web Usage Mining \and Web Robot Detection}
\end{abstract}

\section{Introduction}
The web has become ingrained in our lives, influencing our daily activities and  preserving the web through web archives is more important than before. With over 686 billion web pages archived \cite{IAholdings} dating back to 1996, the Internet Archive (IA) is the largest and oldest of the web archives. The Wayback Machine, which can replay past versions of websites, is a public service provided by IA. Arquivo.pt \cite{billionarch,archsearcharc} has been archiving millions of files from the Internet since 1996. Both web archives contain information in a variety of languages and provide public search capabilities for historical content.\looseness=-1

Our study is an extension of a previous study by AlNoamany et al. \cite{jcdl13:patterns} that examined access patterns for robots and humans in web archives based on a web server log sample from 2012 from the Wayback Machine. By using several heuristics including browsing speed, image to HTML ratio, requests for  robots.txt, and User-Agent strings to differentiate between robot and human sessions, AlNoamany et al. determined that in the IA access logs in 2012, humans were outnumbered by robots 10:1 in terms of sessions, 5:4 in terms of raw HTTP accesses, and 4:1 in terms of megabytes transferred. The four web archive user access patterns defined in the previous study are \textbf{Dip} (single access), \textbf{Slide} (the same page at different archive times), \textbf{Dive} (different pages at roughly the same archive time), and \textbf{Skim} (lists of what pages are archived, i.e., TimeMaps).

We revisit the work of AlNoamany et al. by examining user accesses to web archives using three different datasets from anonymized server access logs: 2012 Wayback Machine (\textbf{IA2012}), 2019 Wayback Machine (\textbf{IA2019}), and 2019 Arquivo.pt (\textbf{PT2019}). Using these datasets, we identify human and robot access, identify important web archive access patterns, and discover the temporal preference for web archive access. We add to the previous study's criteria for distinguishing robots from humans by making a few adjustments. These heuristics will be discussed in detail in Section \ref{bot_identification}. 

The following are the primary contributions of our study:
\begin{enumerate}

\item We used a full-day's worth of three web archive access logs datasets (IA2012, IA2019, PT2019) to distinguish between human and robot access. The total number of robots detected in IA2012 is greater than IA2019 (21\% more in requests and 18\% more in sessions). Robots account for 98\% of requests (97\% of sessions) in PT2019.

\item We looked at different access patterns exhibited by web archive users (humans and robots). We found out that the robots are almost entirely limited to Dip and Skim in IA2012, but exhibit all the established patterns and their combinations in IA2019. 

\item We explored human and robot users' temporal preferences for web archive content. The majority of requests were for mementos that were near to the date-time of each access log dataset, suggesting a preference for the archived content in the recent past.
\end{enumerate}

In this paper, we are attempting to understand who accesses the web archives. To be clear, we are not making any value judgments about robots because we recognize that not all bots are bad. For example, there are services like Internet Archive Scholar \cite{IAscholar}, ArchiveReady \cite{Archivability}, TMVis \cite{mabe_visualizing_2020}, and MemGator \cite{alam2016memgator} which are built on top of web archives that benefit users.  

\section{Background and Related Work}
\label{background}
Extracting useful data from web server logs and analyzing user navigation activity is referred to as web usage mining \cite{webusage_srivastava,varnagar2013web,mughal2018data}. Numerous studies have been conducted for analyzing different web usage mining techniques as well as to identify user access patterns on the Internet \cite{thesis_Cooley,mobasher2005web,liu2011web}. Web usage mining is used to increase the personalization of web-based applications \cite{pierrakos2003web,mobasher2000integrating}. Mobasher et al. \cite{mobasher2000automatic} developed an automatic personalization technique using multiple web usage mining approaches. Web usage mining is also applied in user profiling \cite{grcar2004user,castellano2007web}, web marketing initiatives \cite{berendt2001measuring}, and enhancing learning management systems \cite{zaiane2001web,zaiane2001web}. 

In this work, we look at web archive server access logs and perform web usage mining in the context of web archives. There has been past work in how users utilize and behave in web archives \cite{gomes2013search,costa2013query,hockx2014access,gomes2014importance,costa2011characterizing,ijdl:tpdl13:yasmin}, including the 2013 study \cite{jcdl13:patterns} that we revisit. Web archives maintain their web server access logs as plain text files that record each request to the web archive. Most HTTP servers use the standard Common Log Format or the extended Combined Log Format to record their server access logs \cite{commonlog}. An example access log entry from Arquivo.pt web archive is shown in Figure~\ref{fig:samplelog}. A single log entry consists of the IP address of the client, user identity, authenticated user’s ID, date and time, HTTP method, request path, HTTP version, HTTP status code, and size of the response in bytes, referrer, and User-Agent (left to right). The client IP address is anonymized in the access log datasets for privacy reasons. Alam has implemented an HTTP access log parser \cite{logparser}, with exclusive features for web archive access logs, which can be used to process such web archive access logs. 

Web servers usually operate on the Hypertext Transfer Protocol (HTTP)~\cite{httprfc}. Web clients (such as a web browser or web crawler) make HTTP requests to web servers using a set of defined methods, such as GET, HEAD, POST, etc. to interact with resources~\cite{methodsupport}. Web servers respond using a set of defined HTTP status codes, headers, and payload (if any).


\begin{figure*}
\noindent\fbox{%
    \parbox{\textwidth}{%
        128.82.7.3 - - [07/Jul/2019:04:44:14 +0100] ``GET/wayback/20091223043049/ht
        tp://www.cs.odu.edu/ HTTP/1.1'' 200 9593 ``-'' ``Mozilla/5.0 (X11; Ubuntu; Linux x86\_64; rv:48.0) Gecko/20100101 Firefox/48.0''
    }%
}
\caption{A sample access log entry from the PT2019 dataset (Fields: IP address of the client, user identity, authenticated user’s ID, date and
time, HTTP method, request path, HTTP version, HTTP status code, size of the response in bytes, referrer, and User-Agent)} \looseness=-1
\label{fig:samplelog}
\end{figure*}


The goal of web archives is to capture and preserve original web resources (URI-Rs). Each capture, or memento (URI-M), is a version of a URI-R that comes from a fixed moment in time (Memento-Datetime). The list of mementos for a particular URI-R is called a TimeMap (URI-T). All of these notions are outlined in the Memento Protocol \cite{mementoprotocol}.

AlNoamany et al.'s previous work \cite{jcdl13:patterns} in 2013 set the groundwork for this study. In addition to their analysis of the prevalence of robot and human users in the Internet Archive, they also proposed a set of basic user access patterns for users of web archives:

\begin{itemize}
    \item[] \textbf{Dip} - The user accesses only one URI (URI-M or URI-T).
    \item[] \textbf{Slide} - The user accesses the same URI-R at different Memento-Datetimes.
    \item[] \textbf{Dive} - The user accesses different URI-Rs at nearly the same Memento-Datetime (i.e., dives deeply into a memento by browsing links of URI-Ms).
    \item[] \textbf{Skim} - The user accesses different TimeMaps (URI-T).
\end{itemize}

In a separate study, AlNoamany et al. looked into the Wayback Machine's access logs to understand who created links to URI-Ms and why \cite{tpdl13:wayback,ijdl:tpdl13:yasmin}.  They found that web archives were used to visit pages no longer on the live web (as opposed to prior versions of pages still on the web), and much of the traffic came from sites like Wikipedia.

Alam et al. \cite{archivalvoids_sawood} describe archival voids or portions of URI spaces that are not present in a web archive. They created multiple Archival Void profiles using Arquivo.pt access logs, and while doing so, they identified and reported access patterns, status code distributions, and issues such as Soft-404 (when a web server responds with an HTTP \texttt{200 OK} status code for pages that are actually error pages \cite{soft404}) through Arquivo.pt server logs. While their research is very similar to ours, the mentioned access patterns differ from ours.  Their study looks at which users are accessing the archive and what they request, whereas we explain how a user (robot or human) might traverse through an archive.

\section{Methodology}

\begin{figure}[t]
\centering
\includegraphics[width=\textwidth]{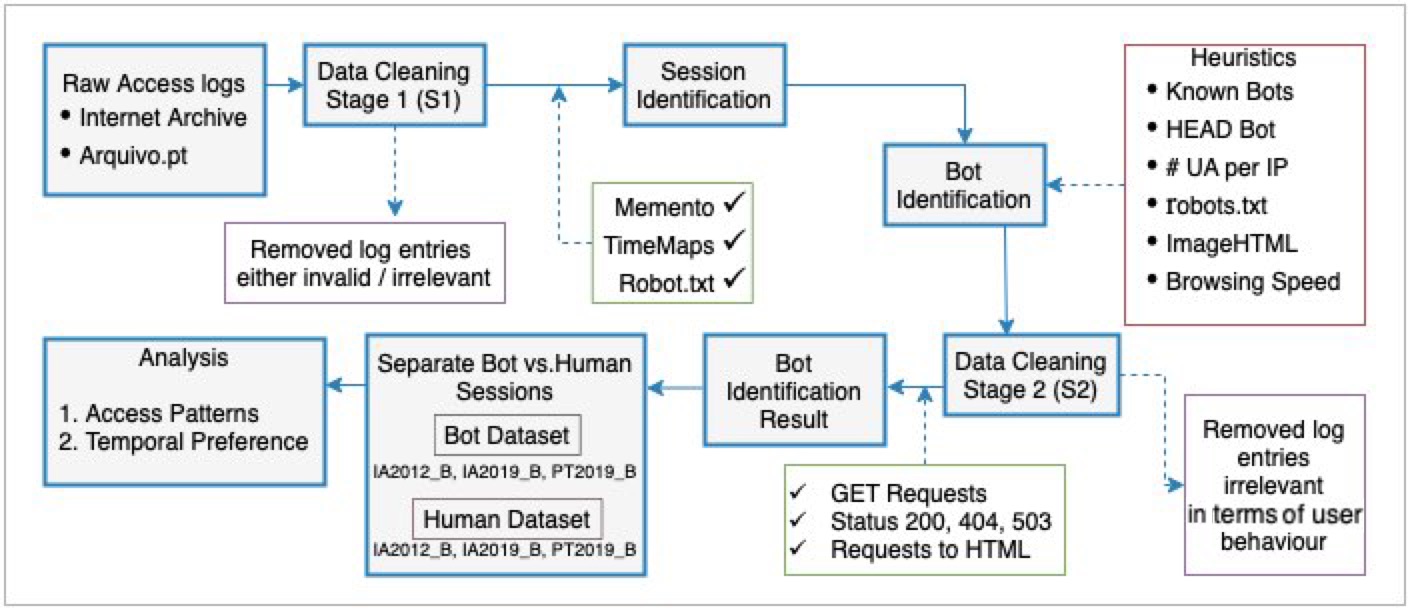}
\caption{A chart illustrating the phases in our analytical procedure}
\label{fig:model}
\end{figure}

In this work, we leverage cleaned access logs after pre-processing raw access logs to identify user sessions, detect robots, assess distinct access patterns used by web archive visitors, and finally check for any temporal preferences in user accesses. The steps of our analysis are shown in Figure~\ref{fig:model}. The code \cite{source_code} and visualizations \cite{observable_notebook} are published and each step is explained in detail in this section. 

\subsection{Dataset}
In this study, we are using three \textbf{full-day} access log datasets from two different web archives: February 2, 2012 access logs from the Internet Archive (IA2012) and February 7, 2019 access logs from Internet Archive (IA2019) and Arquivo.pt (PT2019). We chose the first Thursday of February for our datasets to align with the prior analysis performed on a much smaller sample (2 million requests representing about 30 minutes) from the Wayback access logs from February 2, 2012 \cite{jcdl13:patterns}. 

The characteristics of the raw datasets are listed in Table~\ref{tab:featuretable}. We show the frequency of HTTP request methods and HTTP response codes, among other features. HTTP GET is the most prevalent request method (>98\%) present in all three datasets, while the HTTP HEAD method accounts for less than 1\% of requests. 

Due to the practice of web archives redirecting from the requested  Memento-Datetime to the nearest available memento, IA2012 and IA2019 have numerous 3xx requests. About 20\% of requests are 3xx in PT2019, due to the same behavior. IA2019 has the highest number of requests to embedded resources (63\%) followed by IA2012 (44\%) whereas PT2019 has only 20\%. From IA2012 to IA2019, the number of requests with a null referrer field has decreased by more than half. 
There is an increase in self-identified robots (SI robots) from IA2012 to IA2019. The percentage of SI robots in PT2019 is as twice that in IA2019. We used some of these features (HEAD requests, embedded resources, and SI robots) in the bot identification process (covered in Section~\ref{bot_identification}).

\begin{table}
\centering
\caption{Features for each dataset: February 2, 2012 from IA (IA2012), February 7, 2019 from IA (IA2019), and February 7, 2019 from Arquivo.pt (PT2019).}
\label{tab:featuretable}
\begin{tabular}{l | r r r }
\hline
\multirow{2}{*}{\textbf{Feature}} & \multicolumn{1}{c}{\textbf{IA2012}} & \multicolumn{1}{c}{\textbf{IA2019}} & \multicolumn{1}{c}{\textbf{PT2019}} \\
& \multicolumn{1}{c}{\textbf{ February 2, 2012 }} & \multicolumn{1}{c}{\textbf{ February 7, 2019 }} & \multicolumn{1}{c}{\textbf{ February 7, 2019 }} \\
\hline
No. of Requests & 99,173,542 \textit{(100.00\%)} & 308,194,916 \textit{(100.00\%)} & 1,046,855 \textit{(100.00\%)}\\ 
GET & 97,987,295 \textit{(98.80\%)} & 304,125,661 \textit{(98.68\%)} & 1,025,132 \textit{(97.92\%)}\\
HEAD & 1,109,810 \textit{(1.12\%)} & 2,578,735 \textit{(0.84\%)} & 14,330 \textit{(1.37\%)}\\
PROPFIND & 2,092 \textit{(0.00\%)} & 27,896 \textit{(0.01\%)} & 0 \textit{(0.00\%)}\\
POST & 32,557 \textit{(0.03\%)} & 1,368,941 \textit{(0.44\%)} & 222 \textit{(0.02\%)}\\
OPTIONS & 1,925 \textit{(0.00\%)} & 7,982 \textit{(0.00\%)} & 0 \textit{(0.00\%)}\\ \hline
Status Code 2xx & 32,460,590 \textit{(32.73\%)} & 148,742,768 \textit{(48.26\%)} & 272,467 \textit{(26.03\%)}\\
Status Code 3xx & 52,131,835 \textit{(52.57\%)} & 131,729,104 \textit{(42.74\%)} & 211,709 \textit{(20.22\%)}\\
Status Code 4xx & 11,614,387 \textit{(11.71\%)} & 27,099,599 \textit{(8.79\%)} & 560,913 \textit{(53.58\%)}\\
Status Code 5xx & 2,964,146 \textit{(2.99\%)} & 614,502 \textit{(0.20\%)} & 1,764 \textit{(0.17\%)}\\ \hline
Embedded Resources & 43,260,926 \textit{(43.62\%)} & 195,287,060 \textit{(63.36\%)} & 205,976 \textit{(19.68\%)}\\
Null Referrer & 47,625,026 \textit{(48.02\%)} & 60,935,472 \textit{(19.77\%)} & 265,515 \textit{(25.36\%)}\\
SI Robots & 8,867 \textit{(0.01\%)} & 476,367 \textit{(0.15\%)} & 3,602 \textit{(0.34\%)}\\
\hline
\end{tabular}
\end{table}

\subsection{Data Cleaning}
\label{datacleaning}
An overview of our data cleaning process is shown in Figure~\ref{fig:model}. In the Stage 1 data cleaning (S1), we removed the log entries that were either invalid or irrelevant to the analysis. We only kept legitimate requests to web archive content (mementos and TimeMaps) and requests to the web archive's robots.txt at the end of S1 data cleaning. The robots.txt requests were preserved since they will be utilized as a bot detection heuristic later on in our process. 

After S1 data cleaning, we identified user sessions in each of our three datasets (Section~\ref{session_identification}) and conducted bot identification (Section~\ref{bot_identification}).
Stage 2 data cleaning (S2) takes place only after the requests were flagged as human or robot. Our study's ultimate goal was to detect user access patterns of robots and humans in our datasets, and to do so, we must ensure that the refined datasets only included requests that a user would make. As a result, in S2 we purged log items that were unrelated in terms of user behavior. This includes the browser's automatic requests for embedded resources, any requests using a method other than HTTP GET, and requests generating responses with status codes other than 200, 404, and 503. Several of these requests, including embedded resources and HEAD requests, were necessary during the bot detection phase. Thus, we had to follow a two-step data cleaning approach.

Table~\ref{tab:data_cleaning} shows the number of requests for each dataset after each cleaning stage. The percentages are based on the raw dataset's initial number of requests. PT2019 had a higher percentage of requests remaining after S2 compared to IA2012 and IA2019. This could be related to the raw dataset's low percentage of embedded resources (20\%) in the PT2019 dataset (Table~\ref{tab:featuretable}).


\begin{table}
\centering
\caption{The number of requests in each of the three datasets (IA2012, IA2019, and PT2019): Initial raw data, after stage 1 cleaning, and after stage 2 cleaning.}\label{tab:data_cleaning}
\begin{tabular}{l | r r r }
\hline
\multirow{1}{*}{\textbf{ Dataset }} & \multicolumn{1}{c}{\textbf{\hspace{0.3cm}Raw Dataset}} & \multicolumn{1}{c}{\textbf{\hspace{0.6cm}Stage 1 Cleaning}} & \multicolumn{1}{c}{\textbf{\hspace{0.6cm}Stage 2 Cleaning}} \\
\hline
\multirow{1}{*}{ IA2012} & 99,173,542 & 84,512,394 \textit{(85.22\%)} & 18,432,398 \textit{(18.58\%)}\\
\multirow{1}{*}{ IA2019} & 308,194,916 & 237,901,926 \textit{(77.19\%)} & 35,015,776 \textit{(11.36\%)}\\
\multirow{1}{*}{ PT2019} & 1,046,855 & 904,515 \textit{(86.40\%)} & 604,762 \textit{(57.77\%)}\\
\hline
\end{tabular}
\end{table}

\subsection{Session Identification}
\label{session_identification}

After S1 data cleaning, the next phase in our study was session identification (Figure~\ref{fig:model}). A session can be defined as a set of interactions by a particular user with the web server within a given time frame. We split the requests into different user sessions after S1 data cleaning. First, we sorted all of the requests by IP and User-Agent, then identified the user sessions based on a 10-minute timeout threshold similar to the prior study's process~\cite{jcdl13:patterns}. That is, if the interval between two consecutive requests with the same IP and User-Agent is longer than 10 minutes, the second request is considered as the start of the next session for that user. 

\subsection{Bot identification} 
\label{bot_identification}
As the next step in our process, we employed a heuristic-based strategy to identify robot requests (Figure \ref{fig:model}). We incorporated a few new
adjustments to the original heuristics used in prior work \cite{jcdl13:patterns} to improve the performance of the robot detection. The following sub-sections will go through each heuristic in detail. The real-world examples for each heuristic taken from the web archive access logs is shown in the appendices. \looseness=-1

\subsubsection{Known bots}
\label{sub:Kb}

We created a list of User-Agents that are known to be used by bots. We first constructed $UA_l$, a list of all User-Agent strings from our three datasets. From this list, we compiled $UA_m$ by filtering for  User-Agent strings that contained robot keywords, such as ``bot'', ``crawler'', ``spider'', etc. We compiled a separate bot User-Agent list $UA_d$ by running our full list $UA_l$ through DeviceDetector \cite{devicedetector}, a parser that filters on known bot User-Agent strings. Our final list \cite{knownbot_list} of bot User-Agents $UA_{K_b}$ was constructed by combining $UA_d$ with our keyword set $UA_m$. Any request with a User-Agent found in $UA_{K_b}$ was classified as a robot. Appendix~\ref{app:Kb} provides a real-world example where the “bot” keyword is available on the User-Agent itself.  

\subsubsection{Type of HTTP request method}
\label{sub:HEAD}

Web browsers, which are assumed to be humans, send GET requests for web pages. As a result, we used HEAD requests as an indication of robot behavior. If the request made is a HEAD request, it is considered a robot request, and the session to which it belongs is counted as a robot session. Appendix~\ref{app:HEAD} provides a real-world example where HEAD requests are made.    

\subsubsection{Number of User-Agent per IP (UA/IP)}
\label{sub:UAIP}
There are robots that repeatedly change their User-Agent (UA) between requests to avoid being detected.  The previous study \cite{jcdl13:patterns} found that a threshold of 20 UAs per IP was effective in distinguishing robots from humans. This allows for some human requests behind a proxy or NAT that may have the same IP address but different User-Agents, representing different users sharing a single IP. As discussed in Section~\ref{session_identification}, we sorted the access logs from the three datasets based on IP first and then User-Agent. We marked any requests from IPs that update their User-Agent field more than 20 times as robots. Appendix~\ref{app:UAIP} provides a real-world example where the IP address is changed for each request.  

\subsubsection{Requests to {robots.txt} file}
\label{sub:robotstxt}
A robots.txt \cite{robotstxt,robotstxt:blog} file contains information on how to crawl pages on a website. It helps web crawlers control their actions so that they do not overburden the web server or crawl web pages that are not intended for public viewing. As a result, a request for the robots.txt file can be considered an indication of a robot request. We identified any user who made a request for robots.txt (including query strings) as a robot. Appendix~\ref{app:robotstxt} provides a real-world example where requests are made to the robots.txt file.  

\subsubsection{Browsing Speed (BS)}
\label{sub:BS}
We used browsing speed as a criterion to distinguish robots from humans. Robots can navigate the web far faster than humans. Castellano et al. \cite{castellano2007lodap} found that a human would only make a maximum of one request for a new web page every two seconds. Similar to the previous study \cite{jcdl13:patterns}, we classified any session with a browsing speed faster than one HTML request every two seconds (or, $BS >= 0.5$ requests per second) as a robot. Appendix~\ref{app:BS} provides a real-world example where we can see so many requests in a single second which is unusual for human behavior.

\subsubsection{Image-to-HTML Ratio (IH)}
\label{sub:IH}
Robots tend to retrieve only HTML pages, therefore requests for images can be regarded as a sign of a human user. A ratio of 1:10 images to HTML was proposed by Stassopoulou and Dikaiakos \cite{STASSOPOULOU2009265} and used in the prior study \cite{jcdl13:patterns} as a  threshold for distinguishing robots from humans. We flagged a session requesting less than one image file for every 10 HTML files as a robot session. IH was found to have the largest effect in detecting robots in the prior study's dataset, and this holds true for our three datasets as well. Appendix~\ref{app:IH} provides a real-world example where a session is marked as a robot using the IH ratio. \\

We used the aforementioned heuristics on our three datasets to classify each request as human or robot. If a request/session has been marked as a robot at least by one of the heuristics, we have classified it as a robot. After bot identification but before reporting the final results, we performed S2 as described in Section~\ref{datacleaning}.

\section{Results and Analysis}

    In order to investigate the data further after S2 data cleaning, we divided the dataset into two subsets, human sessions, and bot sessions. For each dataset, we used these two subsets to determine user access patterns and compare them to robot access patterns. Finally, we conducted a temporal analysis of the requests in both subsets for each dataset.
    
    \subsection{Robots vs. Humans}
    \label{botsvshumans} 
    
    Table~\ref{Results_bot} reports the number of detected robots for each dataset based on the total number of sessions and the total number of requests. 
    We counted the number of requests classified as robots based on each heuristic independently (as mentioned earlier, the heuristics are not mutually exclusive, so these numbers across a column do not need to add to exactly 100\%). The final row in the table represents the total number of sessions and requests that are marked as robots after applying all the heuristics together.  
    
    \begin{table}
    \centering
    \caption{Bot identification results based on the total number of sessions and the total number of requests for each dataset: IA2012, IA2019, and PT2019 (the header for each column displays the total number of sessions and requests). The heuristics are not mutually exclusive.}
\label{Results_bot}
\begin{tabular}{l | r r r r r r}
\hline
\multirow{3}{*}   & \multicolumn{2}{c}{\textbf{IA2012}}      & \multicolumn{2}{c}{\textbf{IA2019}}      & \multicolumn{2}{c}{\textbf{PT2019}}     \\
                                       & \textbf{Sessions}  & \textbf{Requests}   & \textbf{Sessions}  & \textbf{Requests}   & \textbf{Sessions}  & \textbf{Requests}  \\
                                       \textbf{Heuristics} & \textbf{1,527,340} & \textbf{22,302,090} & \textbf{2,658,637} & \textbf{42,868,048} & \textbf{3,680}     & \textbf{613,672}   \\
\hline
\multirow{2}{*}{Known Bots}            & 21,423             & 398,053             & 322,379            & 4,969,187           & 884                & 67,453             \\
                                       & \textit{(1.40\%)}  & \textit{(1.78\%)}   & \textit{(12.13\%)} & \textit{(11.59\%)}  & \textit{(24.02\%)} & \textit{(10.99\%)} \\
\multirow{2}{*}{\#UA per IP}           & 5,050              & 756,801             & 5,475              & 1,442,574           & 3                  & 2,636              \\
                                       & \textit{(0.33\%)}  & \textit{(3.39\%)}   & \textit{(0.21\%)}  & \textit{(3.37\%)}   & \textit{(0.08\%)}  & \textit{(0.43\%)}  \\
\multirow{2}{*}{robots.txt}            & 1,958              & 11,074              & 9,296              & 31,452              & 404                & 4,236              \\
                                       & \textit{(0.13\%)}  & \textit{(0.05\%)}   & \textit{(0.35\%)}  & \textit{(0.07\%)}   & \textit{(10.98\%)} & \textit{(0.69\%)}  \\
\multirow{2}{*}{IH Ratio}            & 1,327,896          & 19,893,394          & 1,746,989          & 24,056,112          & 2,916              & 589,363            \\
                                       & \textit{(86.94\%)} & \textit{(89.20\%)}  & \textit{(65.71\%)} & \textit{(56.12\%)}  & \textit{(79.24\%)} & \textit{(96.04\%)} \\
\multirow{2}{*}{Browsing Speed}        & 237,271            & 4,563,851           & 514,878            & 21,176,163          & 1,694              & 162,068            \\
                                       & \textit{(15.53\%)} & \textit{(20.46\%)}  & \textit{(19.37\%)} & \textit{(49.40\%)}  & \textit{(46.03\%)} & \textit{(26.41\%)} \\
\hline
\multirow{2}{*}{\textbf{Total Robots}} & \textbf{1,340,318} & \textbf{20,281,301} & \textbf{1,854,282} & \textbf{29,968,059} & \textbf{3,584}     & \textbf{603,654}   \\
                                       & \textit{(87.76\%)} & \textit{(90.94\%)}  & \textit{(69.75\%)} & \textit{(69.91\%)}  & \textit{(97.39\%)} & \textit{(98.37\%)} \\
\hline
\end{tabular}
\end{table}

The Image-to-HTML ratio (IH) had the largest effect on detecting robots across all three datasets. The impact of IH was $\approx$85-90\% in IA2012 but only around $\approx{55-65\%}$ in IA2019. In PT2019, $\approx{80-96\%}$ of robots were detected using the IH ratio, which is higher compared to IA2019. In PT2019, we were able to detect almost all the robots through this one heuristic, IH. We found that $\approx{90\%}$ of requests were robots in IA2012, $\approx{70\%}$ of requests were robots in IA2019, and $\approx{98\%}$ of requests were robots in PT2019. 

The reason for this increase in human sessions in 2019 than in 2012 could be the increase in awareness of web archives among human users. In addition, headless browsers, such as Headless Chromium \cite{HeadlessChromium}, PhantomJS \cite{PhantomJS}, and Selenium \cite{selenium}, that provide automated web page control have become popular in recent years. Their functionality simulates a more human-like behavior that may not be caught easily by bot detection techniques. For instance, applications like the work of Ayala \cite{reyes2022correspondence} and tools like the oldweb.today \cite{dshr:oldweb,newoldweb}, DSA Toolkit \cite{jones_dissertation_nonanon,dsa_code4lib_2022}, TMVis \cite{mabe_visualizing_2020}, and Memento-Damage service \cite{mementodamage} that replicate human behavior make things challenging for detection algorithms. Between IA2019 and PT2019, PT2019 has $\approx{30\%}$ more robots present. Based on our PT2019 dataset, only 2\% of all requests coming into the Arquivo.pt are potential human requests.

\subsection{Discovering Access Patterns}
\label{access_pattern}

Upon distinguishing robots from humans, we divided all three of our datasets into human and bot subdatasets (IA2012\_H, IA2012\_B, IA2019\_H, IA2019\_B, PT2019\_H, PT2019\_B). We used these datasets to identify different access patterns that are followed by both human and robot sessions. Upon distinguishing robots from humans, we divided all three of our datasets into human and bot subdatasets (IA2012\_H, IA2012\_B, IA2019\_H, IA2019\_B, PT2019\_H, PT2019\_B). As introduced in Section~\ref{background}, there were four different user access patterns established by AlNoamany et al. \cite{jcdl13:patterns}. We looked into each of these patterns and identified their prevalence in our three datasets. We discovered the prevalence of sessions that followed each of the four patterns (Dip, Dive, Slide, Skim), as well as sessions that followed a hybrid of those patterns (``Dive and Slide'',  ``Dive and Skim'', ``Skim and Slide'', and ``Dive, Slide, and Skim''). We categorized requests that do not fall into any pattern as \textbf{Unknown}. 

\begin{figure}
\centering
\includegraphics[width=\textwidth]{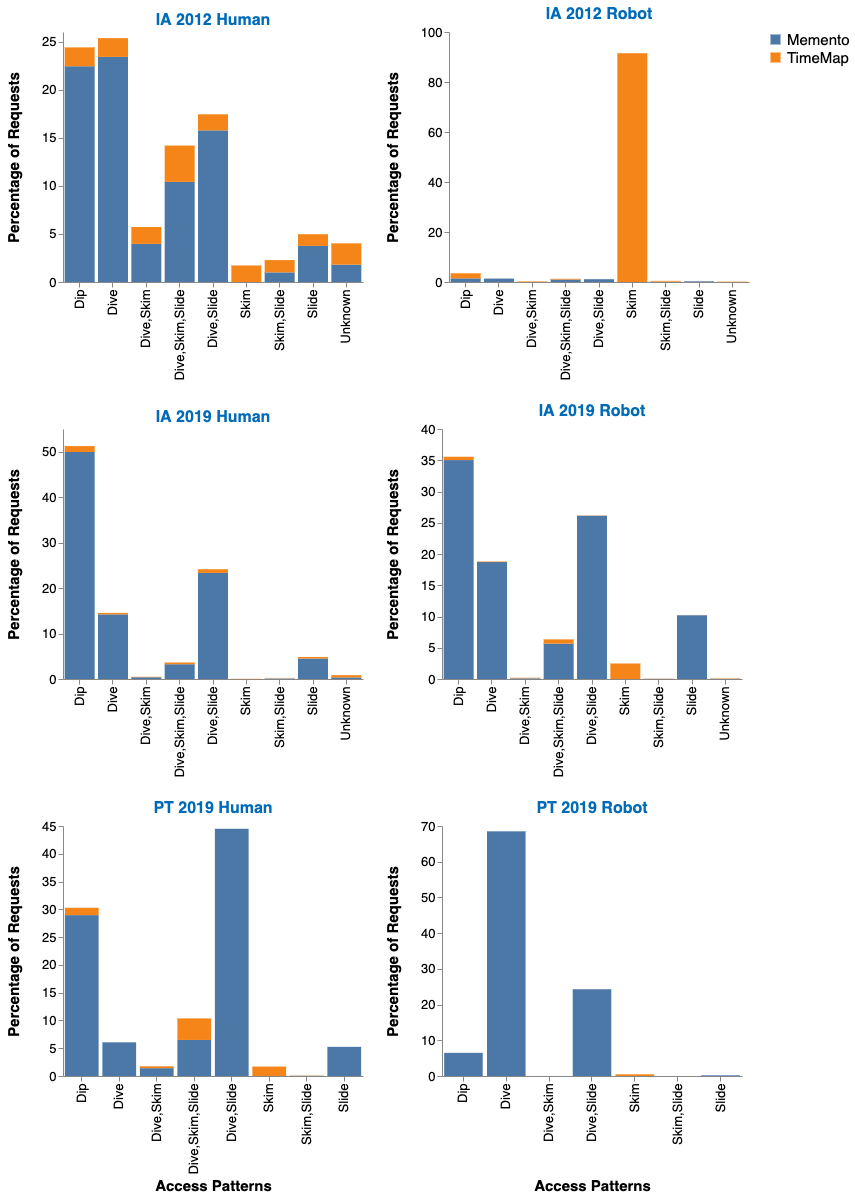}
\caption{Access patterns of robots and humans in our subdatasets (IA2012\_H, IA2012\_B, IA2019\_H, IA2019\_B, PT2019\_H, PT2019\_B). The color of the stacked bar distinguishes between requests for mementos (URI-Ms) and TimeMaps (URI-Ts). Note that the y axes in the charts are not the same.}
\label{fig:access_patterns}
\end{figure}

Figure~\ref{fig:access_patterns} shows a chart for each subdataset. The horizontal (x) axis represents the different patterns or a hybrid of patterns and the vertical (y) axis represents the percentage of the number of requests. The percentages are based on the total number of requests for each subdataset. According to AlNoamany et al.'s findings based on the IA2012 dataset, \textbf{Dips} were the most common pattern in both human and robot sessions. However in our IA2012 dataset (full-day), \textbf{Dive} and \textbf{Dip} account for about the same percentage of human sessions, although \textbf{Skim} is the most common pattern among robot sessions. \textbf{Dip} is the most common pattern in IA2019, followed by \textbf{Dive, Slide} for both human and robot sessions. The human \textbf{Dips} have more than doubled from IA2012 (22\%) to IA2019 (50\%) indicating that more humans are accessing web archives to access a single URI-M or URI-T. There are a high number of robot \textbf{Skims} in IA2012 compared to IA2019. In IA2012 robot sessions, it is over 90\% \textbf{Skims}. We could see that the long-running robot sessions that request URI-Ts account for most of the \textbf{Skim} percentage. In contrast to IA2019, PT2019 humans exhibit a higher percentage of \textbf{Dive} and \textbf{Slide} (45\%) than \textbf{Dips} (29\%). Even in robot sessions, \textbf{Dive} (70\%) and \textbf{Dive} and \textbf{Slide} (24\%) percentage is higher than \textbf{Dip} (6\%). \looseness=-1

In IA2012, robots almost always access TimeMaps (95\%) and humans access mementos  (82\%). However, in IA2019, humans and robots almost always access mementos (96\%), whereas only 4\% of those accesses are to TimeMaps. When looking at the hybrid patterns, PT2019 bot sessions only have a maximum of two patterns while the rest have a small percentage of all three patterns (\textbf{Dive, Skim, and Slide}). For each dataset in IA, there is a very small percentage of requests (4.22\% in IA2019, 0.97\% in IA2012) that do not belong to any of the patterns. We were able to identify all the different patterns in the PT2019 dataset. The percentage of human requests falling under the \textbf{Unknown} category in IA2012 (4.02\%) is higher compared to the IA2012 robot requests (0.2\%), IA2019 human requests (0.85\%), and IA2019 robot requests (0.12\%).\looseness=-1

\subsection{Identifying Temporal Preferences}
\label{temporal_preference}

\begin{figure}[t]
\centering
\includegraphics[width=\textwidth]{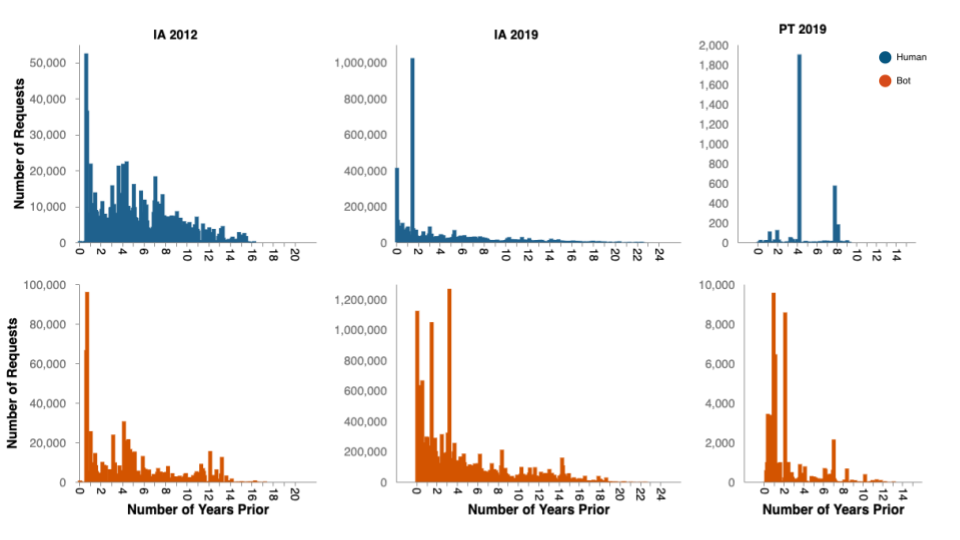}
\caption{Temporal preference of bots and humans in IA2012, IA2019, PT2019 datasets.}
\label{fig:temporal_pref}
\end{figure}

We also explored the requested Memento-Datetime in our subdatasets to see if there was any temporal preference by web archive users. Figure~\ref{fig:temporal_pref} illustrates the temporal preference of robots and humans in our datasets. The x-axis represents the number of years prior, meaning the number of years passed relative to the datetime of the access logs (e.g., for IA2012, 2 years prior is 2010) and the y-axis represents the number of requests. Note that the y-axis in each chart is different. 

It is evident that the majority of the requests are for mementos that are close to the datetime of each access log sample and gradually diminish as we go further back in time. There is no significant difference in temporal preference in IA2012 and IA2019. IA2019 humans, IA2019 bots, and PT2019 bots exhibit the same trend however, it is difficult to see a trend in PT2019 humans due to the fewer number of humans in the dataset. For PT2019 humans, there is a spike around 4-5 years prior which implies PT human accesses were mostly for mementos around 2015-2016. There is an advantage to knowing the temporal preferences of web archive users. Web archives can prioritize or store data in memory for the most recent years to speed up disk access.

\section{Future Work}

AlNoamany et al. \cite{jcdl13:patterns} observed four different user access patterns in 2013. In our datasets combined, 0.35\% of requests were outside of any of these patterns or their combinations. One may look into if the percentage of requests that fell into the \textbf{Unknown} category have any other generally applicable patterns, or if they are completely random. 

The overall number of robots identified in IA2019 is much lower than in IA2012. We would like to repeat this study on more distinct full-day datasets to see if the reduction in robots is a general behavior from 2012 to 2019 or specific to the day we chose. Additionally, the IH \cite{STASSOPOULOU2009265} and BS thresholds \cite{castellano2007lodap} in our bot identification heuristics are based on the behavior of conventional web servers; however, it remains to be determined if the same thresholds apply to web archival replay systems, as the dynamics of web archival replay systems differ (e.g., the Wayback Machine is typically slower than a typical web server). 

\section{Conclusions}

We used a full-day access logs sample of Internet Archive's (IA) Wayback Machine from 2012 and 2019, as well as Arquivo.pt's from 2019, to distinguish between robot and human users in web archives. In IA2012, the overall number of robots discovered was higher than in IA2019 (21\% more in requests and 18\%t more in sessions). We discovered that robot accesses account for 98\% of requests (97\% of sessions) based on 2019 server logs from Arquivo.pt. We also discovered that in IA2012, the robots were almost exclusively limited to Dip and Skim, but that in IA2019, they exhibit all of the patterns and their combinations. Regardless of whether it is a robot or a human user, the majority of requests were for mementos that are close to the date-time of each access log dataset, demonstrating a preference for the recent past.

\section{Acknowledgements}

We thank Mark Graham, director of the Wayback Machine, for sharing access log data from the Wayback Machine at the Internet Archive. We are grateful to Daniel Gomes and Fernando Melo of Arquivo.pt for sharing access log data from the Arquivo.pt web archive with us.

\clearpage
\bibliographystyle{splncs04}
\bibliography{refs}

\newpage
\appendix 

\section{Known Bots}\label{app:Kb}

This heuristic makes use of a list of User-Agents that are recognized as being used by bots. We first constructed a list of all User-Agent strings from our three datasets. From this list, we compiled a list by filtering for  User-Agent strings that contained robot keywords, such as ``bot'', ``crawler'', ``spider'', etc. Our GitHub repository hosts the comprehensive list of known bots \cite{knownbot_list} that was created. Below is an example where ``Twitterbot/1.0'' is the User-Agent. Section~\ref{sub:Kb} discusses this heuristic in more detail. 

\begin{lstlisting}
199.16.157.100_0_0 - - [07/Jul/2019:14:00:01 +0100] "GET /robots.txt HTTP/1.1" 200 1414 "-" "@Twitterbot/1.0@"

199.16.157.100_0_0 - - [07/Jul/2019:14:00:01 +0100] "GET /robots.txt HTTP/1.1" 200 1414 "-" "@Twitterbot/1.0@"

199.16.157.100_0_0 - - [07/Jul/2019:14:00:02 +0100] "HEAD /wayback/20170625001353/http://www.fabricadochocolate.com HTTP/1.1" 200 - "-" "@Twitterbot/1.0@" 

199.16.157.100_0_0 - - [07/Jul/2019:14:00:02 +0100] "HEAD /wayback/20170625001353/http://www.fabricadochocolate.com HTTP/1.1" 200 - "-" "@Twitterbot/1.0@"

199.16.157.100_0_0 - - [07/Jul/2019:14:00:05 +0100] "HEAD /wayback/20170625001353/http://www.fabricadochocolate.com/ HTTP/1.1" 200 - "-" "@Twitterbot/1.0@"

199.16.157.100_0_0 - - [07/Jul/2019:14:00:05 +0100] "HEAD /wayback/20170625001353/http://www.fabricadochocolate.com/ HTTP/1.1" 200 - "-" "@Twitterbot/1.0@"

199.16.157.100_0_0 - - [07/Jul/2019:14:00:07 +0100] "HEAD /wayback/20170625001353/http://www.fabricadochocolate.com/ HTTP/1.1" 200 - "-" "@Twitterbot/1.0@"

199.16.157.100_0_0 - - [07/Jul/2019:14:00:07 +0100] "HEAD /wayback/20170625001353/http://www.fabricadochocolate.com/ HTTP/1.1" 200 - "-" "@Twitterbot/1.0@"
\end{lstlisting}

\newpage
\section{Type of HTTP Request Method}\label{app:HEAD}

We used HEAD requests as an indication of robot behavior. If the request made is a HEAD request, it is considered a robot request, and the session to which it belongs is counted as a robot session. Below is an example where HTTP HEAD requests are made to different mementos. The User-Agent is "Twitterbot" in these request logs, which is another indication that they are robot requests.  Section~\ref{sub:HEAD} discusses this heuristic in more detail. 

\begin{lstlisting}
199.16.157.100_0_0 - - [07/Jul/2019:14:00:02 +0100] "@HEAD@ /wayback/20170625001353/http://www.fabricadochocolate.com HTTP/1.1" 200 - "-" "Twitterbot/1.0"

199.16.157.100_0_0 - - [07/Jul/2019:14:00:02 +0100] "@HEAD@ /wayback/20170625001353/http://www.fabricadochocolate.com HTTP/1.1" 200 - "-" "Twitterbot/1.0"

199.16.157.100_0_0 - - [07/Jul/2019:14:00:05 +0100] "@HEAD@ /wayback/20170625001353/http://www.fabricadochocolate.com/ HTTP/1.1" 200 - "-" "Twitterbot/1.0"

199.16.157.100_0_0 - - [07/Jul/2019:14:00:05 +0100] "@HEAD@ /wayback/20170625001353/http://www.fabricadochocolate.com/ HTTP/1.1" 200 - "-" "Twitterbot/1.0"

199.16.157.100_0_0 - - [07/Jul/2019:14:00:07 +0100] "@HEAD@ /wayback/20170625001353/http://www.fabricadochocolate.com/ HTTP/1.1" 200 - "-" "Twitterbot/1.0"

199.16.157.100_0_0 - - [07/Jul/2019:14:00:07 +0100] "@HEAD@ /wayback/20170625001353/http://www.fabricadochocolate.com/ HTTP/1.1" 200 - "-" "Twitterbot/1.0"
\end{lstlisting}

\newpage
\section{Number of User-Agents per IP (UA/IP)}\label{app:UAIP}

There are robots that repeatedly change their User-Agent (UA) between requests to avoid being detected. We marked any requests from IPs that update their User-Agent field more than 20 times as robots. Below is an example where the IP address is changed for each request. Section~\ref{sub:UAIP} discusses this heuristic in more detail.

\begin{lstlisting}
0.77.87.100 - - [02/Feb/2012:03:46:54 +0000] "POST http://web.archive.org/web/20070211155651/http://212.227.83.57/cproc.aspx HTTP/1.0" 302 0 "http://www.vbleisure.co.uk/guest\_book.html" "@Mozilla/4.0 (compatible; MSIE 5.5; Windows NT 4.0)@"

0.77.87.100 - - [02/Feb/2012:04:06:29 +0000] "POST http://web.archive.org/web/20070211155651/http://212.227.83.57/cproc.aspx HTTP/1.0" 302 - "http://www.vbleisure.co.uk/guest\_book.html" "@Mozilla/4.0 (compatible; MSIE 6.0; Windows NT 5.1; SV1; .NET CLR 1.1.4322)@"

0.77.87.100 - - [02/Feb/2012:05:09:30 +0000] "POST http://web.archive.org/web/20070211155651/http://212.227.83.57/cproc.aspx HTTP/1.0" 302 - "http://www.vbleisure.co.uk/guest\_book.html" "@Mozilla/4.0 (compatible; MSIE 6.0; Windows NT 5.0)@"	

0.77.87.100 - - [02/Feb/2012:07:59:43 +0000] "POST http://web.archive.org/web/20070501120942/http://www.ibcmemorial.org.way\_back\\_stub/formmailer.php HTTP/1.0" 302 0 "http://ibcmemorial.org/sign-guestbook.html" "@Mozilla/4.0 (compatible; MSIE 6.0; Windows NT 5.1; ru) Opera 8.50@"

. . . 

. . .  

0.77.87.100 - - [02/Feb/2012:22:04:57 +0000] "POST http://web.archive.org/web/20070501120942/http://www.ibcmemorial.org.way\_back\_stub/formmailer.php HTTP/1.0" 503 - "http://ibcmemorial.org/sign-guestbook.html" "@Mozilla/4.0 (compatible; MSIE 6.0; Windows 98; Win 9x 4.90; Creative)@"

0.77.87.100 - - [02/Feb/2012:22:08:02 +0000] "POST http://web.archive.org/web/20070501120942/http://www.ibcmemorial.org.way\_back\_stub/formmailer.php HTTP/1.0" 503 - "http://ibcmemorial.org/sign-guestbook.html" "@Mozilla/4.0 (compatible; MSIE 6.0; Windows NT 5.1; .NET CLR 1.0.3705)@"

0.77.87.100 - - [02/Feb/2012:23:40:31 +0000] "POST http://web.archive.org/web/20070501120942/http://www.ibcmemorial.org.way\_back\_stub/formmailer.php HTTP/1.0" 503 - "http://ibcmemorial.org/sign-guestbook.html" "@Mozilla/4.0 (compatible; MSIE 6.0; Windows NT 5.1; en) Opera 9.0@"

0.77.87.100 - - [02/Feb/2012:23:40:32 +0000] "POST http://web.archive.org/web/20070501120942/http://www.ibcmemorial.org.way\_back\_stub/formmailer.php HTTP/1.0" 503 - "http://ibcmemorial.org/sign-guestbook.html" "@Mozilla/4.0 (compatible; MSIE 6.0; Windows NT 5.1; MRA 4.6 (build 01425))@"

0.77.87.100 - - [02/Feb/2012:23:59:34 +0000] "POST http://web.archive.org/web/20070501120942/http://www.ibcmemorial.org.way\_back\_stub/formmailer.php HTTP/1.0" 503 - "http://ibcmemorial.org/sign-guestbook.html" "@Opera/7.60 (Windows NT 5.2; U)  [en] (IBM EVV/3.0/EAK01AG9/LE)@"
\end{lstlisting}

\newpage
\section{Requests to {robots.txt}}\label{app:robotstxt}
A robots.txt file contains information on how to crawl pages on a website. As a result, a request for the robots.txt file can be considered an indication of a robot request. The requests made to the web archives' robots.txt file are demonstrated in the examples that follow. Section~\ref{sub:robotstxt} discusses this heuristic in more detail.

\begin{lstlisting}
0.139.100.213_2_2 - - [02/Feb/2012:17:03:22 +0000] "@GET http://web.archive.org/robots.txt@ HTTP/1.1" 200 125 "-" "RSS Scout 0.9.2"

0.139.100.213_2_2 - - [02/Feb/2012:17:06:30 +0000] "GET http://web.archive.org/web/*/http://c00lbookmarks.com/story.php?title=best-door-blinds-inside HTTP/1.1" 302 0 "-" "RSS Scout 0.9.2"

0.139.100.213_2_2 - - [02/Feb/2012:17:06:32 +0000] "GET http://wayback.archive.org/web/*/http://c00lbookmarks.com/story.php?title=best-door-blinds-inside HTTP/1.1" 404 2409 "http://web.archive.org/web/*/http://c00lbookmarks.com/story.php?title=best-door-blinds-inside" "RSS Scout 0.9.2"

0.139.100.213_2_2 - - [02/Feb/2012:17:07:38 +0000] "@GET http://web.archive.org/robots.txt@ HTTP/1.1" 200 125 "-" "RSS Scout 0.9.2"

0.139.100.213_2_2 - - [02/Feb/2012:17:10:44 +0000] "GET http://web.archive.org/web/*/http://www.goloco.org/users/D5EWwXI HTTP/1.1" 302 0 "-" "RSS Scout 0.9.2"

0.139.100.213_2_2 - - [02/Feb/2012:17:10:45 +0000] "GET http://wayback.archive.org/web/*/http://www.goloco.org/users/D5EWwXI HTTP/1.1" 404 2385 "http://web.archive.org/web/*/http://www.goloco.org/users/D5EWwXI" "RSS Scout 0.9.2"

0.139.100.213_2_2 - - [02/Feb/2012:17:14:50 +0000] "@GET http://web.archive.org/robots.txt@ HTTP/1.1" 200 125 "-" "RSS Scout 0.9.2"

0.139.100.213_2_2 - - [02/Feb/2012:17:19:54 +0000] "@GET http://web.archive.org/robots.txt@ HTTP/1.1" 200 125 "-" "RSS Scout 0.9.2"
\end{lstlisting}

\newpage
\section{Browsing Speed (BS)}\label{app:BS}
We used browsing speed as a criterion to distinguish robots from humans. Robots can navigate the web far faster than humans. It was found that a human would only make a maximum of one request for a new web page every two seconds. We classified any session with a browsing speed faster than one HTML request every two seconds (or, $BS >= 0.5$ requests per second) as a robot. The example below demonstrates how a single IP made several requests in a single second. Section~\ref{sub:BS} discusses this heuristic in more detail.

\begin{lstlisting}
0.0.100.100_0_0 web.archive.org - [@07/Feb/2019:00:46:30 +0000@] "GET /web/20190205174131/https://connect.facebook.net/signals/config/225699104785488?v=2.8.40&r=stable HTTP/1.1" 200

0.0.100.100_0_0 web.archive.org - [@07/Feb/2019:00:46:30 +0000@] "GET /web/20190207001831/https://fonts.googleapis.com/css?family=Lato:100,100i,300,300i,400,400i,700,700i,900,900i&subset=latin-ext HTTP/1.1" 200

0.0.100.100_0_0 web.archive.org - [@07/Feb/2019:00:46:30 +0000@] "GET /web/20190207001831/https://fonts.googleapis.com/css?family=Lato:100,100i,300,300i,400,400i,700,700i,900,900i&subset=latin-ext HTTP/1.1" 200

0.0.100.100_0_0 web.archive.org - [@07/Feb/2019:00:46:30 +0000@] "GET /web/20190207001831/https://fonts.googleapis.com/css?family=Lato:100,100i,300,300i,400,400i,700,700i,900,900i&subset=latin-ext HTTP/1.1" 200

0.0.100.100_0_0 web.archive.org - [@07/Feb/2019:00:46:30 +0000@] "GET /web/20190207001831/https://fonts.googleapis.com/css?family=Lato:100,100i,300,300i,400,400i,700,700i,900,900i&subset=latin-ext HTTP/1.1" 200

0.0.100.100_0_0 web.archive.org - [@07/Feb/2019:00:46:30 +0000@] "GET /web/20190207004025/https://connect.facebook.net/en_US/fbevents.js HTTP/1.1" 200

0.0.100.100_0_0 web.archive.org - [@07/Feb/2019:00:46:30 +0000@] "GET /web/20190207004025/https://connect.facebook.net/signals/config/225699104785488?v=2.8.40&r=stable HTTP/1.1" 302

0.0.100.100_0_0 web.archive.org - [@07/Feb/2019:00:46:30 +0000@] "GET /web/20190207004026/https://embed.tawk.to/59cc85aec28eca75e4622ccd/default HTTP/1.1" 200

0.0.100.100_0_0 web.archive.org - [@07/Feb/2019:00:46:30 +0000@] "GET /web/20190207004026/https://embed.tawk.to/59cc85aec28eca75e4622ccd/default HTTP/1.1" 200

0.0.100.100_0_0 web.archive.org - [@07/Feb/2019:00:46:30 +0000@] "GET /web/20190207004026/https://fonts.googleapis.com/css?family=Lato:100,100i,300,300i,400,400i,700,700i,900,900i&subset=latin-ext HTTP/1.1" 302
\end{lstlisting}

\newpage
\section{Image-to-HTML Ratio (IH)}\label{app:IH}

Robots tend to retrieve only HTML pages, therefore requests for images can be regarded as a sign of a human user. We flagged a session requesting less than one image file for every 10 HTML files as a robot session. The below is an example where only requests for HTML files are made without any images or other embedded resources. Section~\ref{sub:IH} discusses this heuristic in more detail.

\begin{lstlisting}
0.0.122.100_1_0 web.archive.org - [07/Feb/2019:16:55:22 +0000] "@GET /web/*/http://maestro.haarp.alaska.edu/@ HTTP/2.0" 200 9002 "https://archive.org/search.php?query=http%3A%2F%2Fmaestro.haarp.alaska.edu%2F" "Mozilla/5.0 (Windows NT 10.0; Win64; x64) AppleWebKit/537.36 (KHTML, like Gecko) Chrome/70.0.3538.102 Safari/537.36 OPR/57.0.3098.116" 0.192 MISS 0.192 "@text/html@; charset=utf-8" - "-" "-" "wwwb-app31" "-"

0.0.122.100_1_0 web.archive.org - [07/Feb/2019:16:56:15 +0000] "@GET /web/20130304102141/http://maestro.haarp.alaska.edu/@ HTTP/2.0" 404 0 "https://web.archive.org/web/20130715000000*/http://maestro.haarp.alaska.edu/" "Mozilla/5.0 (Windows NT 10.0; Win64; x64) AppleWebKit/537.36 (KHTML, like Gecko) Chrome/70.0.3538.102 Safari/537.36 OPR/57.0.3098.116" 10.859 MISS 10.856 "@text/html@; charset=utf-8" - "-" "-" "wwwb-app104" "-"

0.0.122.100_1_0 web.archive.org - [07/Feb/2019:16:56:15 +0000] "@GET /web/20130304102141/http://maestro.haarp.alaska.edu/@ HTTP/2.0" 404 0 "https://web.archive.org/web/20130715000000*/http://maestro.haarp.alaska.edu/" "Mozilla/5.0 (Windows NT 10.0; Win64; x64) AppleWebKit/537.36 (KHTML, like Gecko) Chrome/70.0.3538.102 Safari/537.36 OPR/57.0.3098.116" 10.926 MISS 10.928 "@text/html@; charset=utf-8" - "-" "-" "wwwb-app58" "-"

0.0.122.100_1_0 web.archive.org - [07/Feb/2019:16:56:15 +0000] "@GET /web/20130304102141/http://maestro.haarp.alaska.edu/@ HTTP/2.0" 404 0 "https://web.archive.org/web/20130715000000*/http://maestro.haarp.alaska.edu/" "Mozilla/5.0 (Windows NT 10.0; Win64; x64) AppleWebKit/537.36 (KHTML, like Gecko) Chrome/70.0.3538.102 Safari/537.36 OPR/57.0.3098.116" 11.453 MISS 11.456 "@text/html@; charset=utf-8" - "-" "-" "wwwb-app57" "-"

. . . 

. . .

0.0.122.100_1_0 web.archive.org - [07/Feb/2019:16:56:23 +0000] "@GET /web/20130304102141/http://maestro.haarp.alaska.edu/@ HTTP/2.0" 404 8274 "https://web.archive.org/web/20130715000000*/http://maestro.haarp.alaska.edu/" "Mozilla/5.0 (Windows NT 10.0; Win64; x64) AppleWebKit/537.36 (KHTML, like Gecko) Chrome/70.0.3538.102 Safari/537.36 OPR/57.0.3098.116" 0.000 HIT - "@text/html@; charset=utf-8" - "-" "-" "wwwb-app43" "-"

0.0.122.100_1_0 web.archive.org - [07/Feb/2019:16:56:23 +0000] "@GET /web/20130304102141/http://maestro.haarp.alaska.edu/@ HTTP/2.0" 404 8274 "https://web.archive.org/web/20130715000000*/http://maestro.haarp.alaska.edu/" "Mozilla/5.0 (Windows NT 10.0; Win64; x64) AppleWebKit/537.36 (KHTML, like Gecko) Chrome/70.0.3538.102 Safari/537.36 OPR/57.0.3098.116" 0.227 MISS 0.224 "@text/html@; charset=utf-8" - "-" "-" "wwwb-app43" "-"

0.0.122.100_1_0 web.archive.org - [07/Feb/2019:16:56:29 +0000] "@GET /web/*/http://maestro.haarp.alaska.edu/*@ HTTP/2.0" 200 8341 "https://web.archive.org/web/20130304102141/http://maestro.haarp.alaska.edu/" "Mozilla/5.0 (Windows NT 10.0; Win64; x64) AppleWebKit/537.36 (KHTML, like Gecko) Chrome/70.0.3538.102 Safari/537.36 OPR/57.0.3098.116" 0.087 MISS 0.088 "@text/html@; charset=utf-8" - "-" "-" "wwwb-app57" "-"

\end{lstlisting}

\end{document}